\documentclass[sigconf,nonacm]{acmart}
\usepackage{cleveref}
\usepackage{booktabs}
\usepackage{pifont}
\usepackage{xspace}
\usepackage{float}
\usepackage{listings}

\lstdefinelanguage{Markdown}{
    morekeywords={\#}, 
    sensitive=false,
    morecomment=[l]{\#},
    morestring=[b]",
    morestring=[b]',
    moredelim=[is][\color{blue}\bfseries]{**}{**},
    moredelim=[is][\color{teal}\ttfamily]{`}{`},
    alsoletter={\#}
}

\AtBeginDocument{%
  }

\newcommand{\systemname}{AlterAtlas\xspace}
\newcommand{\subsub}[1]{\noindent\textbf{#1}}

\acmISBN{978-1-4503-XXXX-X/2018/06}

\begin{document}

\title{AlterAtlas: Shifting Travel Planning from AI Generation to Validation via Persona-Driven Simulations}

\author{William Huang}
\orcid{0000-0001-7651-2190}
\affiliation{%
  \institution{University of California, Los Angeles}
  \city{Los Angeles}
  \state{CA}
  \country{USA}
}
\email{william.huang@ucla.edu}

\author{Ruofei Du}
\orcid{}
\affiliation{%
  \institution{Google}
  \city{San Francisco}
  \state{CA}
  \country{USA}
}
\email{me@duruofei.com}

\author{Yang Zhang}
\orcid{0000-0003-2472-6968}
\affiliation{%
  \institution{University of California, Los Angeles}
  \city{Los Angeles}
  \state{CA}
  \country{USA}
}
\email{yangzhang@ucla.edu}

\renewcommand{\shortauthors}{Huang et al.}
\begin{abstract}
Travel planning requires balancing interacting goals and constraints across time and space. Current AI travel tools provide limited support for encoding these constraints and understanding how generated travel plans may fail users. We present \systemname, an interactive travel planning system that supports high-fidelity itinerary validation and revision through persona-based simulations grounded in geospatial information. \systemname models travelers as editable personas, generates candidate itineraries from prioritized places of interest, and simulates how different personas would experience each plan. Simulations expose route-level tradeoffs, temporal user states (e.g., fatigue, hunger), and mismatches between plans and user preferences to allow users to iteratively refine both itineraries and user personas. An expert evaluation of 51 paired itineraries demonstrates that simulation-guided revisions significantly improve plan-persona alignment. Furthermore, a within-subjects study (N=11) reveals that \systemname empowers users to uncover hidden constraints, fluidly compare alternatives, and build trust in their final plans. Our results suggest that simulation-based validation is a powerful, transparent interaction layer for AI-assisted travel planning.
\end{abstract}

\begin{teaserfigure}
  \includegraphics[width=\textwidth]{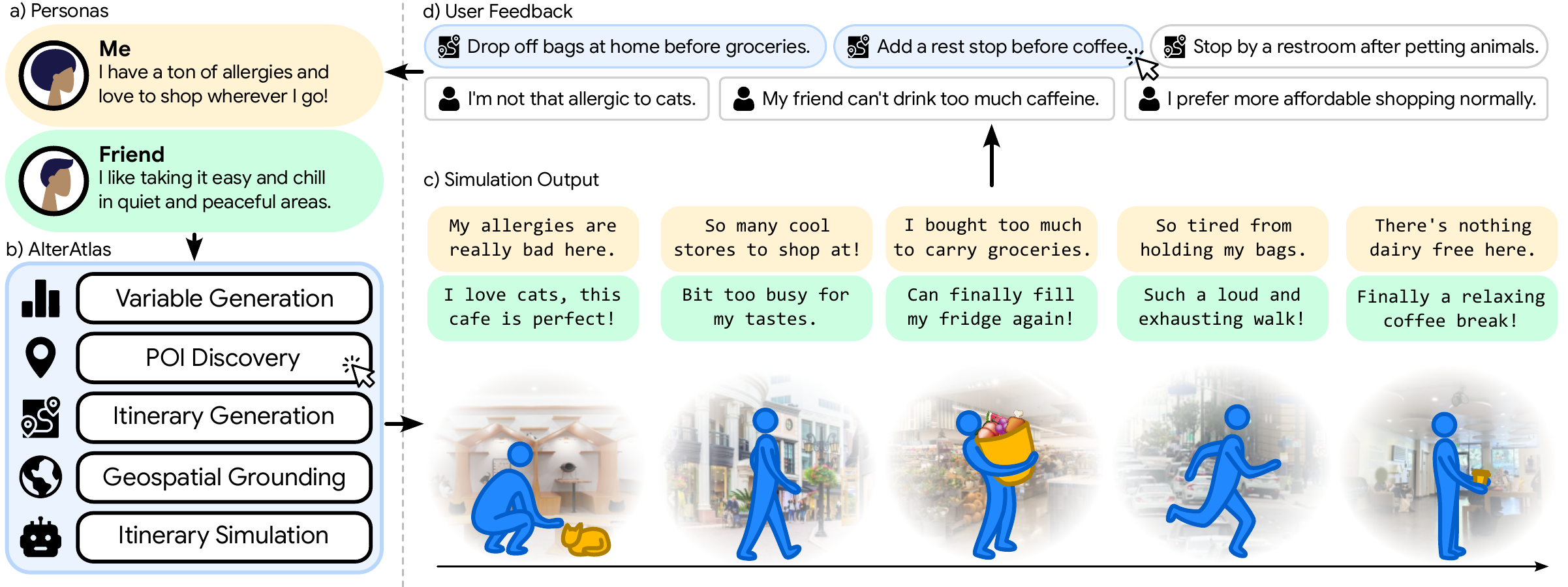}
  \caption{\systemname shifts AI travel planning from one-shot generation to interactive validation by simulating how different travelers experience a route through a four-stage workflow. a) Users define a set of personas representing themselves or travel companions. b) \systemname uses these personas alongside geospatial data to generate candidate routes and step-by-step simulated experiences. c) Users use simulations to validate travel plans, shifting the focus away from one-shot itinerary construction to iterative validation. d) Simulations allow users to identify key improvements to both their travel plans and the user persona assumptions. Arrows indicate the workflow progression.}
  \Description{Diagram of the \systemname workflow for interactive travel planning. Users define personas with different preferences, such as ``Me'' with allergies and interest in shopping and ``Friend'' who prefers quiet areas. \systemname then performs variable generation, POI discovery, itinerary generation, geospatial grounding, and itinerary simulation. Given user constraints and edits, the system simulates how each persona experiences the route, producing reactions such as crowdedness, convenience, fatigue, shopping appeal, and availability of dairy-free options, so users can refine the itinerary.}
  \label{fig:teaser}
\end{teaserfigure}

\maketitle

\section{Introduction}

Travel planning is a multi-constraint spatial and temporal optimization problem \cite{susani.stewartCaseBasedApproachUnderstanding1999}. Travelers must balance goals such as must-see places, appointments, and desired experiences against interacting, dynamic constraints such as time, mobility, fatigue, hunger, and accessibility. As these constraints rarely operate independently, adding a single stop can fundamentally alter the walking burden, meal timing, crowd exposure, or overall route feasibility later on in the day \cite{zhuVirtualCommunityUsers2023}. Consequently, travel planning is inherently a tedious process of repeated revision and replanning.

Recent LLM-based travel assistants enable itinerary generation from high-level prompt descriptions, with emerging systems \cite{chenTravelAgentAIAssistant2024, otakiTravelItineraryRecommendation2025, volchekChatGPTTravelItinerary2024, zhangEnhancingTravelPlanning2024, wangHowDoesChatGPTgenerated2025} increasingly supporting trip recommendation, route creation, and itinerary editing. However, these systems predominantly center planning around one-shot generation of a finished plan, assuming a ``correct'' travel plan while leaving complex user constraints and potential plan failure modes unconsidered and unexplained. \cite{shenTripTailorRealWorldBenchmark2025, zouResearchTravelRoute2024}. As such, these systems fail to support the core iteration, revision, and contingency identification workflows that characterize how people actually travel plan. 

In this paper, we investigate simulation-based validation as a complementary interaction layer for AI-assisted travel planning. We introduce \systemname, a system that treats itinerary generation not as an endpoint, but as the start of a workflow centered around validation, comparison, and revision. \systemname represents travelers as editable personas with user-defined constraints, leveraging these profiles alongside geospatial data to simulate how users would actually \textit{experience} candidate travel plans. These simulations expose route-level tradeoffs, in-the-moment fluctuations in user states, and potential mismatches between a plan and the people it is intended to support. Using these simulations as a representation and elicitation mechanism for spatial preferences, users can then use these insights to iteratively revise itineraries, tweak persona assumptions, and compare alternatives. Through \systemname, we introduce a novel interaction paradigm for travel planning in which simulations serve as inspectable planning artifacts that externalize possible constraints and support revision.

Our approach is grounded in a formative study (N=7), which showed that travel planning is highly iterative and organized around place-of-interest (POI) prioritization, route comparisons, and mental simulations. \systemname supports this natural workflow by preserving POI prioritization while supporting constraint inspection and revision. We evaluate \systemname through two complementary studies: First, we investigate how simulation-guided revision improves personalized travel plan quality through an expert evaluation of 51 matched initial/updated itinerary pairs. Second, we explore how \systemname fits into and enhances \textit{real-world} planning practices through a within-subjects user study (N=11). Together these studies assess both the quality of generated plans and the interaction value of simulation-based revisions. 
We find that simulation-based revision significantly improves expert-rated plan-persona alignment. Furthermore, participants effectively used \systemname to surface hidden constraints, confidently revise both plans and user personas, compare alternatives, and better understand potential outcomes with real-world travel plans.

Our primary contributions are: 
\begin{enumerate}
    \item Design implications derived from a formative study (N=7) detailing how people plan travel and how existing AI tools fall short in supporting iterative workflows.
    \item A persona-based travel plan simulator that uses editable personas and grounded geospatial data to generate in-situ user perceptions and states in travel plans.
    \item \systemname, an interactive travel planning system that shifts the AI generation from one-shot to simulation-mediated validation and refinement through a novel interaction paradigm.
    \item Findings from an expert-driven study and user study (N=11) demonstrating how \systemname improves generated plan alignment and empowers user-driven iterative workflows.
\end{enumerate}

\section{Related Works}
We situate \systemname within prior literature on LLM-based travel planning, systems for spatial and embodied assessment of physical environments, and simulation-drive HCI systems.

\subsection{LLM-Based Travel Planning}
Recent advances in LLM-based planning \cite{wangSurveyLargeLanguage2024} have enabled a growing ecosystem of AI applications for travel planning \cite{volchekChatGPTTravelItinerary2024, zhangEnhancingTravelPlanning2024, kumamotoExploringRoleLarge2025, zouResearchTravelRoute2024, parkEnhancingEffectivenessGenerative2025}. Prior works have explored this domain through multimodal AI systems \cite{bermejoxiaophotos2025}, browser-based assistance \cite{udandaraoRoamifyDesigningEvaluating2025}, web interaction data \cite{otakiTravelItineraryRecommendation2025}, interactive maps \cite{dengIMAIAInteractiveMaps2025}, recommender agents \cite{singhAutomatedTravelPlanning2024, zhangEnhancingTravelPlanning2024}, and benchmarks \cite{xieTravelPlannerBenchmarkRealWorld2024,huangITIMOLLMempoweredSynthesis2026, shenTripTailorRealWorldBenchmark2025, chengTravelBenchBroaderRealWorld2026}. Recent works focus on personalization, taking individual user needs and preferences into account during the generation process \cite{TripGeniePersonalized, sPersonalizedTravelItinerary2024, aribasTransformingPersonalizedTravel2024, shenTripTailorRealWorldBenchmark2025, gargEnhancingTailoredTravel2025}. Commercial applications similarly emphasize generating or editing complete itineraries from high-level user requests through systems like Layla \cite{MeetLaylaAI}, Booking.com AI Agents \cite{BookingcomEnhancesTravel}, TripGenie \cite{TripGeniePersonalized}, and Gemini for Google Maps \cite{HowWereReimagining2026}. Together, these tools have made travel planning more automated; however, their dominant interaction model still centers around candidate plan generation rather than user understanding \cite{zouResearchTravelRoute2024}.

\systemname builds upon this line of work but shifts the point of interaction. Rather than treating travel planning as a one-shot generation task, \systemname treats itinerary generation as the first stage in a iterative workflow of validation, comparison, and revision. By using simulations as a verification layer and grounding mechanism, \systemname enables users to express goals and constraints more effectively than through prompt authoring alone \cite{zamfirescu-pereiraWhyJohnnyCant2023}.

\subsection{Embodied and Spatial Assessments of Physical Environments}
A second body of work examines how people and systems inspect unfamiliar physical environments. Systems like Project Sidewalk \cite{sahaProjectSidewalkWebbased2019b} and Urban Sidewalks \cite{boltenUrbanSidewalksVisualization2016} support accessibility mapping, remote auditing, and routing using street view imagery and geospatial accessibility data. More recent research has explored richer forms of remote embodied assessment. Geollery supports immersive exploration of geotagged information in shared 3D mirrored worlds \cite{duGeolleryMixedReality2019}. Bring Environments to People uses virtual tours for remote accessibility assessments \cite{chiBringEnvironmentsPeople2023}. Embodied Exploration pushes this further by enabling wheelchair users to assess accessibility through VR-based embodied interactions with digital replicas of physical spaces \cite{peiEmbodiedExplorationFacilitating2023a}. Closest to our approach is Accessibility Scout \cite{huangAccessibilityScoutPersonalized2025a} which uses LLMs and collaborative assessments to generate personalized accessibility scans of built environments and incrementally refine a user model. While these systems demonstrate the value of personalized spatial assessments, they generally evaluate single POIs. Rather, we contend that in real traveling situations, constraints must be taken in context to the overall route where a place suitable in isolation might fail to satisfy user needs when hungry, tired, or stressed. 

Adjacent work on urban-scene perception and multimodal geospatial reasoning takes a more automated approach, using street imagery and vision-language models to infer complex human signals such as urban socioeconomic status \cite{liuCityLensBenchmarkingLarge2025}, human perceptions of safety and class \cite{mushkaniVisionLanguageModelsSee2025}, urban attractiveness \cite{zhouEvaluatingUrbanVisual2025}, and physical measurements of urban clearances \cite{perezStreetscapeAnalysisGenerative2025}. Closest to our work is Generative Traffic Agents \cite{lammerGTAGenerativeTraffic2026} which uses llm-based persona simulations to model urban-scale spatial behaviors. These systems primarily estimate scene-level properties, leaving individual user-facing validations and embodied reasoning relatively unexplored. \systemname unifies these approaches in a new interaction model, where simulations are used as a new data modality to simultaneously balance automated geospatial evaluations with dynamic user planning goals.

\subsection{Simulation-Driven HCI Systems}

Personas have recently gained traction in HCI as data-informed representations of users that support both design reasoning and system evaluation. Shin et al. \cite{shinUnderstandingHumanAIWorkflows2024} found that hybrid human-AI workflows where humans interpret data and LLMs synthesize narratives produce more statistically representative personas. PersonaCraft \cite{jungPersonaCraftLeveragingLanguage2025} and Park et al. \cite{parkGenerativeAgentSimulations2024a} similarly demonstrate that LLMs can transform survey data into coherent personas at scale. More recent works extend personas beyond static artifacts by turning them into interactive systems, enabling conversational interfaces for usability, engagement, and qualitative inquiries \cite{benharrakWriterDefinedAIPersonas2024, kaateWhenPersonasTalk2025, rizwanPersonaBOTBringingCustomer2025}. Azem et al. \cite{azemInterviewingAIGeneratedPersonas} uses persona-based interviews while CloChat \cite{haCloChatUnderstanding2024} shows that customizable AI personas foster richer dialogue. ChoiceMates \cite{parkChoiceMatesSupportingUnfamiliar2026a} similarly uses multiagent persona-based conversations to support unfamiliar online decision making. Simulations have also seen widespread use for evaluating and testing systems at scale. Prior work has used simulated agents to iteratively improve LLM systems \cite{belleAgentsChangeSelfEvolving2025}, model long-term social interactions \cite{parkGenerativeAgentsInteractive2023a}, and societal simulations \cite{piaoAgentSocietyLargeScaleSimulation2025}. Simulation-based evaluations have also been applied to websites and recommendation systems through systems like UXAgent \cite{luUXAgentSystemSimulating2025}, AgentA/B \cite{wangAgentAAutomatedScalable2025}, and Chen et al. \cite{chenWhyAmSeeing2025}. Closest to our work, City Sim \cite{bougieCitySimModelingUrban2025} simulates urban behavior for city-scale policy evaluation. 

\systemname draws from this body of work, extending personas beyond static summaries to editable, route-conditioned user personas that drive prospective simulations of candidate itineraries. Simulations are shown directly to users as interactive planning artifacts, enabling them to inspect hidden constraints, diagnose misalignments, and iteratively revise both plans and personas. 

\section{Formative Study}
\label{sec:formative}
To inform the design of \systemname, we conducted a formative study to understand how people currently plan trips, pain points of existing workflows, and what kind of AI support would be most useful. 

\subsection{Procedure}
We recruited seven frequent travel planners (Age: 23-31; 3M/4F; F1-F7; Minimum Monthly Travel Planners) with experience using generative AI tools for travel planning. Through 30-minute interviews, participants described their current planning workflows, how they evaluate travel plan success, and reflected on AI tools for travel planning. We then showed short examples of potential end-to-end AI travel planner systems and asked participants to brainstorm about their ideal planning tools. Audio recordings and notes were analyzed using codebook thematic analysis \cite{braunReflectingReflexiveThematic2019a} by two researchers. Quotes have been lightly edited for concision, grammar, and anonymity.

\subsection{Findings}
Participants described planning as an iterative balancing process, constantly weighing the enjoyment of adding a new POI against the potential costs like fatigue and hunger. This friction was exacerbated in group contexts, where user needs frequently conflict. Rather than constructing a route from start to finish, participants began by identifying a set of high-priority goals, explored alternatives around those anchors, and mentally simulated how constraints might accumulate across a day. Below, we detail the key findings that directly motivate the design of \systemname.

\subsub{Planning is structured around anchors and priorities.}
All participants (F1-F7) began with identifying \textit{must-go} anchors such as core attractions, reservations, hotels, or appointments. They then searched for intermediate stops that fit around and enabled a more enjoyable experience at these anchors. In this sense, participants viewed itineraries not as flat sequences of stops, but as priority-structured plans in which some destinations were fixed and others remained negotiable. When asked how to evaluate the quality of a plan, all participants responded that a plan can be considered successful if it enables them to efficiently and comfortably reach all these anchors. Intermediate stops were instead framed as supportive choices that fill gaps, satisfy needs, or reduce risks rather than evaluation-defining destinations in their own right. As F5 noted, \textit{``if I have a list of places I want to see and I hit all the places, I'd say [my plan] was pretty successful.'' } This suggests that as long as core user goals are preserved in travel plans, intermediary routing choices and POIs are largely interchangeable and can be abstracted.

\subsub{Users mentally simulate how constraints unfold across plans.}
Participants evaluated routes by imagining how local conditions and personal constraints would interact over time. Examples include: checking whether there would be restrooms at the right times when bringing kids (F7), whether a pedestrian crossing or parking lot would feel stressful (F5), whether a long walk or hill would be tiring (F2), whether drives would be diverse and pretty (F4), or whether meal timings would make sense given shifting hunger and satiety throughout the day (F5). These considerations were often interdependent, where participants noted spending considerable effort attempting to predict how hidden constraints would shape the overall experience. This challenge was further exacerbated in group planning scenarios, where participants (F1, F6, F7) noted that it was even more difficult to keep track of what others would appreciate based on prior travel experiences. Participants therefore wanted more than route generation, requesting assistance in understanding how user states and constraints could be affect the overall experience of the travel plan.

\subsub{Users want iterative refinement and preserved alternatives.}
Participants (F3, F4, F5, F6, F7) treated plans as provisional, saved multiple options, and wanted to adapt them without rebuilding from scratch. Although participants varied in how much automation they wanted, they consistently asked for support that allowed them to remain at a high level like an overall plan summary when desired and drill down into micro-level changes like routing recommendations when necessary. As such, high-level summaries allow for coarse filtering of potential candidate plans and microlevel understanding informs candidate plan refinement. Participants explicitly requested controllable refinement, where they could compare alternatives, update assumptions, and revise parts of the plan without full reconstruction. As F5 put it, for an AI-based travel planner to be successful, \textit{``you need constant user input.''} (F5). Our findings indicate that iterative refinement is enabled through controllable refinement at various levels of granularity, where users should know exactly what changed and how those changes compare to alternatives.

\subsub{Planning builds confidence in future travels.}
Traveling is a high-commitment activity that requires a certain degree of trust and confidence in a plan \cite{shinTravelDecisionDeterminants2022, liuScaleDevelopmentTourist2019}. As such, participants all (F1-F7) noted that verification and trust-building are integral parts of the travel planning process. For many of the participants (F2, F3, F5, F6), the process of researching places and cross-checking route feasibility was how they established trust in an itinerary, following prior work in travel plan trust \cite{agagWhyConsumersTrust2017,  choiOnlineTravelInformation2021, parkEnhancingEffectivenessGenerative2025}. However, current AI tools were noted to abstract too much of the validation process: \textit{``I don't want it to be a loading screen and give me the plan. I want some sort of progression [of reasoning].''} (F5); \textit{``If I do not get enough references I have to do my own homework and I hate that.''} (F1). As such, participants only references trust in the context of the travel plan, where any tools mentioned (e.g., web search, generative AI) always required latter manual plan validation for trust. 

\subsection{Design Considerations}
Based on these insights, we outline four design implications for personalized AI travel planning systems.

\subsub{D1: Support priority-structured planning itineraries.} 
Systems should preserve user-defined priorities and prioritization autonomy by distinguishing between the required and flexible elements of an itinerary. AI support should respect this hierarchy, preserving user-defined anchors while abstracting lower-level routing work. 

\subsub{D2: Externalize route-level user states and constraints.} 
Systems should support modeling of user constraints by representing user states explicitly. Users must be able to inspect and revise these assumptions as their understanding of travel needs change. 

\subsub{D3: Enable progressive refinement and comparison.}
Systems should support lightweight modification and re-evaluation of candidate routes, enabling users to compare and revise plans under changing traveler states while iteratively refining the routes and user personas. 

\subsub{D4: Surface evidence and simulation traces for plan inspection.}
Systems should surface evidence-backed recommendations and simulations that help users inspect and diagnose where a candidate plan misaligns with route conditions, user constraints, or persona assumptions to ground supervised revisions.

\section{\systemname}

\begin{figure}
    \centering
    \includegraphics[width=1\linewidth]{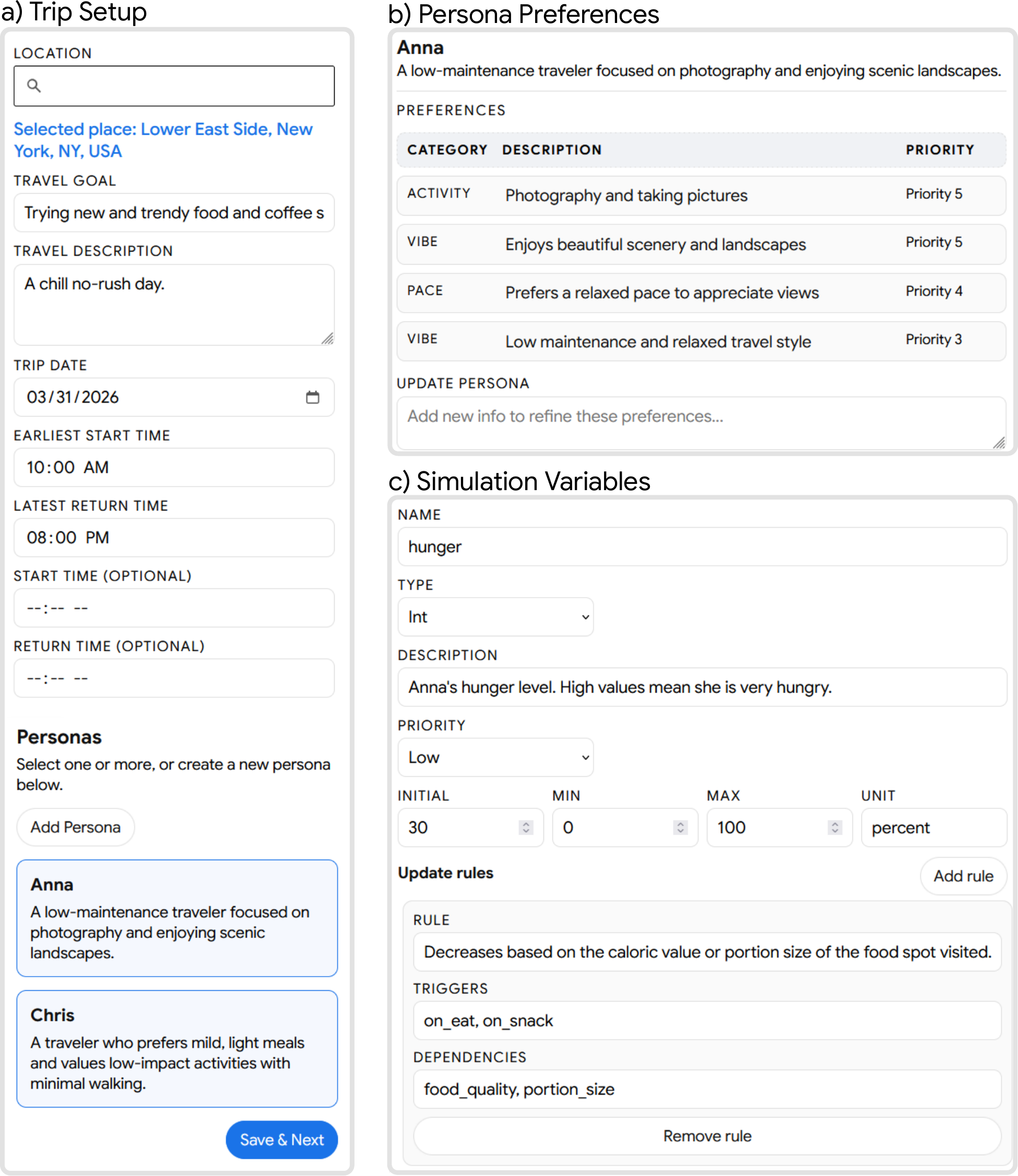}
    \caption{\systemname setup interfaces. a) Trip parameter setup interface. b) Persona preference viewer. c) Persona simulation variable viewer.}
    \Description{a) A trip setup web UI. The page consists of multiple text entry and drop down selections for location, date, and trip descriptions. There is also a selection UI at the bottom to create personas and select existing personas. b) The persona preferences web UI. At the top is a table of preferences that list category, description, and a priority value. Under is a textbox to add updates to the list. c) The simulation variables interface. The page consists of multiple text entries to define various rules and descriptions for new simulation variables including the ability to add a variable number of update rules.
}
    \label{fig:setup-ui}
\end{figure}

\begin{figure}
    \centering
    \includegraphics[width=1\linewidth]{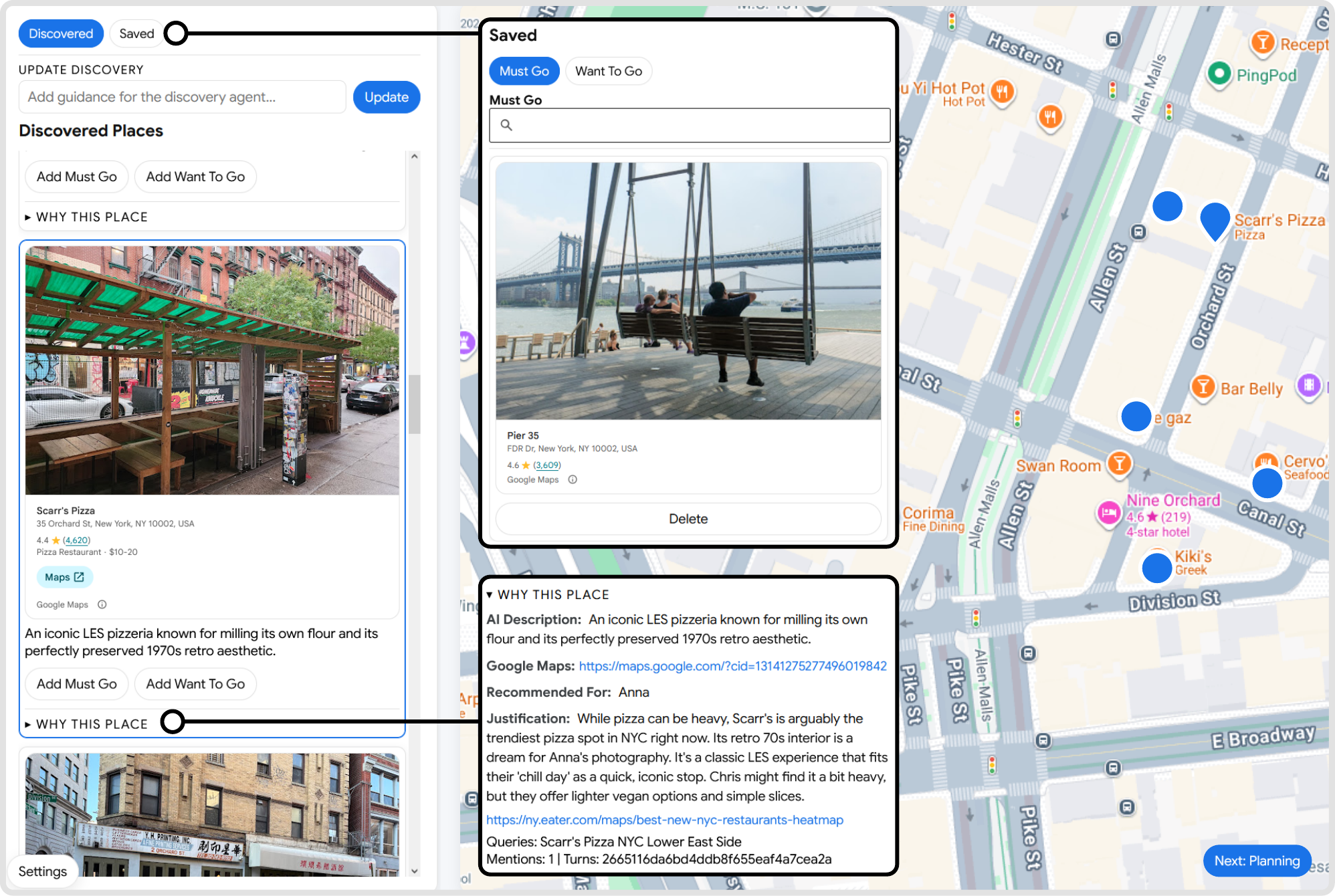}
    \caption{Discovery user interface. Users are shown recommended POIs on a 2D map. Users can group each POI as either ``Must Go'' or ``Want to Go'' which they can view and manually add new POIs through the ``Save'' button. Recommended POIs are shown with justifications for why they were recommended.}
    \label{fig:discovery-ui}
    \Description{Three \systemname discovery interfaces are shown: a map-based POI discovery view, a saved places panel, and a POI justification panel. Users can browse recommended places on a city map, save them as ``Must Go'' or ``Want to Go'', add their own POIs, and inspect textual explanations for why a place was recommended.}
\end{figure}

\begin{figure*}
    \centering
    \includegraphics[width=1\linewidth]{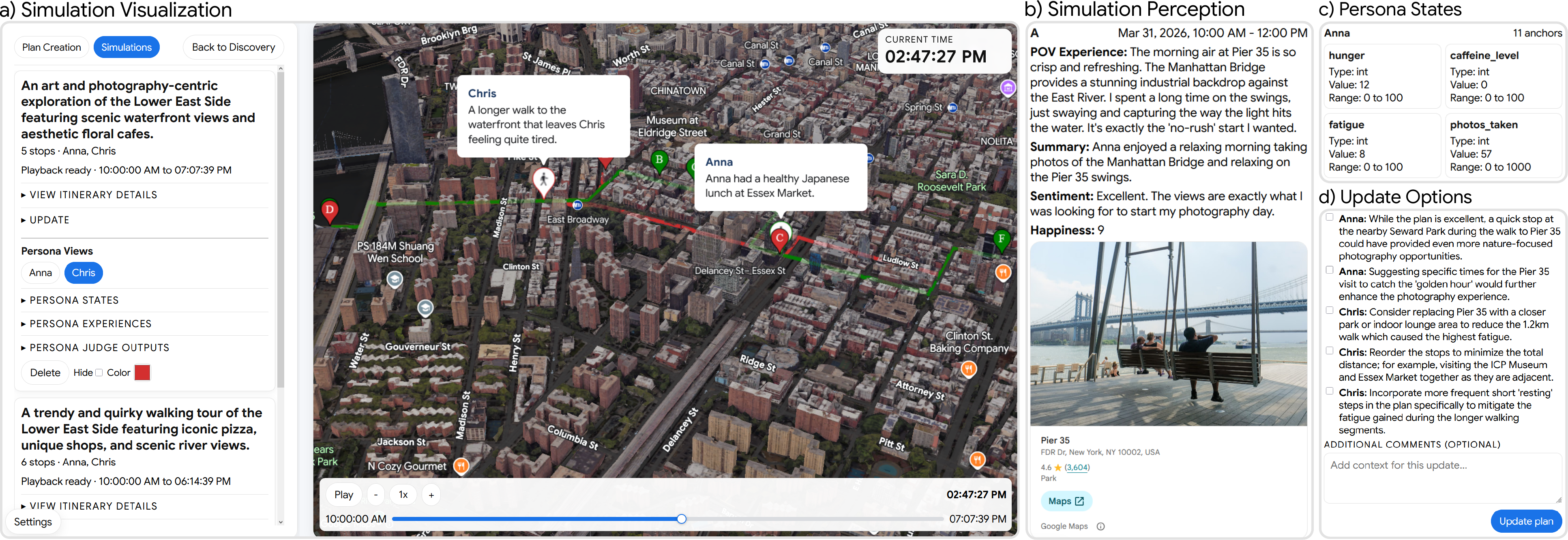}
    \caption{Simulation user interface. a) Users begin by generating travel plans and simulations which are shown in an interactive 3D environment. Users can scrub through the trip time to validate where personas are, how they are feeling, and geospatial data. b) Example simulation perception for a single POI. c) Example persona state. d) Example list of suggested improvements.}
    \Description{a) Simulation viewer web UI. On the left is a sidebar with information on different travel plans. On the right is a 3D map of the area with simulated travel plans plotted. At the bottom is a scrubber with the time. Each travel plan has playhead markers to show where a user would be and a small dialogue box showing what the simulation feels at that moment. b)  Simulation perception web UI. A web box with information on what the simulation experience was, a summary of that experience, general sentiment, and a happiness score. At the bottom is the link to that POI. c) Persona states web UI. A set of boxes which depict the various simulation variables for a persona. Shows the current value of that variable and the range. d) Update options web UI. A list of suggested updates with labels for who the improvement is suggested for. Checkboxes allow users to select suggested improvements or add their own comments through the textbox below.
}
    \label{fig:simulation-ui}
\end{figure*}

Grounded in our formative study findings (\Cref{sec:formative}), we developed \systemname, a travel-planning system that uses persona-based simulations to validate candidate itineraries. To emphasize route-level and micro-level constraint modeling, \systemname focuses on single-day walking-only itineraries. \systemname supports trip planning through a four-stage workflow (\Cref{fig:teaser}): users first define the trip context and personas of all travelers and potential traveler states (\Cref{fig:setup-ui}a); next, \systemname supports place-of-interest (POI) discovery and prioritization by generating grounded recommendations that help distinguish protected anchors from flexible options; it then generates candidate itineraries that satisfy selected POIs and user preferences; finally, users run persona-based simulations on each candidate to examine route-level tradeoffs, compare alternatives, and iteratively revise both plans and personas. 

\subsection{\systemname Walkthrough}
\label{sec:system-walkthrough}
We illustrate \systemname through the following scenario. Anna is planning a day trip to the \textit{Lower East Side, New York City} with friends. She specifies a trip on \textit{March 31, 2026} from \textit{10:00AM--8:00PM} with the goal of \textit{``trying new and trendy food and coffee spots''} and notes she wants a \textit{``a chill no-rush day''} (\Cref{fig:setup-ui}a). She creates a persona for herself, \textit{``Low-maintenance traveler who likes taking pictures and enjoying views''}, and another for her friend Chris, \textit{``Dislikes heavy and spicy foods and doesn't like walking too much''} (\Cref{fig:setup-ui}b). \systemname generates simulation variables for each, such as Anna's \verb|hunger| and Chris's \verb|fatigue| which accumulates faster than normal. Anna views these and adds \verb|caffeine_level| to the Anna persona to reflect her coffee jitters (\Cref{fig:setup-ui}).

In discovery (\Cref{fig:discovery-ui}), Anna reviews suggested POIs, saving several as \textit{must-go} and \textit{want-to-go},  \systemname then generates multiple candidate itineraries and displays them on the 3D map. Anna takes a quick look at each and simulates both personas across all candidates and inspects their traces (\Cref{fig:simulation-ui}ab). One plan keeps Anna happy but includes too many coffee stops while the other drives Chris's fatigue too high due to long walking segments (\Cref{fig:simulation-ui}c). She deletes one candidate and revises another by accepting a suggested rest stop for Chris and replacing some of the coffee stops with lower-caffeine boba stores (\Cref{fig:simulation-ui}d). After rerunning the simulations, she finds that the updated plan better satisfies both personas and saves it. Anna is unsure how tired she will be from work that day, so she adds a new preference to her persona (\textit{``gets sleepy early''}) and simulates a contingency plan (\Cref{fig:setup-ui}b).

\subsection{System Design and Implementation}
We describe the design and implementation of \systemname below.

\subsub{Personas and simulation variables.}
\systemname personalizes planning through editable personas, structured natural-language profiles that describe a traveler's goals, tolerances, and preferences (D2) (\Cref{fig:setup-ui}b). Personas are first initialized through a natural language user description which are then parsed and organized into structured fields across 12 preference categories (travel style, pace, budget, safety, sustainability, accessibility, social, food, vibe, activity, culture, cost) which users can update and add at any time. For each planning run, personas are then converted into a set of typed \textit{simulation variables} representing route-relevant state and constraints that can differ per trip and user through LLM-based schema generation. Variables may be bounded integers, booleans, or enums which each include (1) update rules describing how it changes across itinerary steps and (2) dependencies specifying which route or user-state factors influence it. Users are able to view and modify any simulation variables throughout \systemname (\Cref{fig:setup-ui}c). 

For example, a simulation variable such as \verb|hunger| may be represented as an integer from 0-100 with update rules like ``increase over time spent traveling'' and ``decrease after meals in proportion to meal size'' and with dependencies on \verb|elapsed_time|, \verb|fatigue|, and \verb|distance_walked|. Simulation variables are fully freeform and editable at anytime. \verb|fatigue| and \verb|hunger| being initialized by default.

\subsub{POI discovery and prioritization.}
The discovery interface supports the priority-structured planning pattern identified in our formative study (D1). An LLM-based recommender follows realistic user search flows for POI recommendations: (1) researching interesting areas and neighborhoods, (2) Identifying how they fit with persona preferences, and (3) filtering and structuring POI recommendations. The discovery agent is equipped with Google Places and web search tools to support this. Each recommendation includes a short justification and linked web evidence  (\Cref{fig:discovery-ui}) (D4). Users then classify locations into ``must-go'' and ``want-to-go'' buckets, allowing \systemname to distinguish protected anchors from flexible options before itinerary generation. Users are able to add known POIs or filter through recommended POIs (\Cref{fig:discovery-ui}).

\subsub{Candidate itinerary generation.}
Given prioritized POIs, a consolidation agent generates itineraries that preserves ``must-go'' locations, prioritizes ``want-to-go'' locations when possible, and adds supporting stops as needed through web search and Google Places tooling. Through a mapping tool, the agent can visualize POIs and routes for grounded mapping knowledge. The agent can also perform rerouting through adding intermediate latitude-longitude waypoints. Generated itineraries are shown on an interactive 3D map, where users can inspect routes and request coarse edits like reordering stops, replacing POIs, or shortening segments. We note that without any prioritized POIs, the consolidation agent functions similarly to existing travel planning agents like TravelAgent \cite{chenTravelAgentAIAssistant2024}.

\begin{figure*}
    \centering
    \includegraphics[width=.95\linewidth]{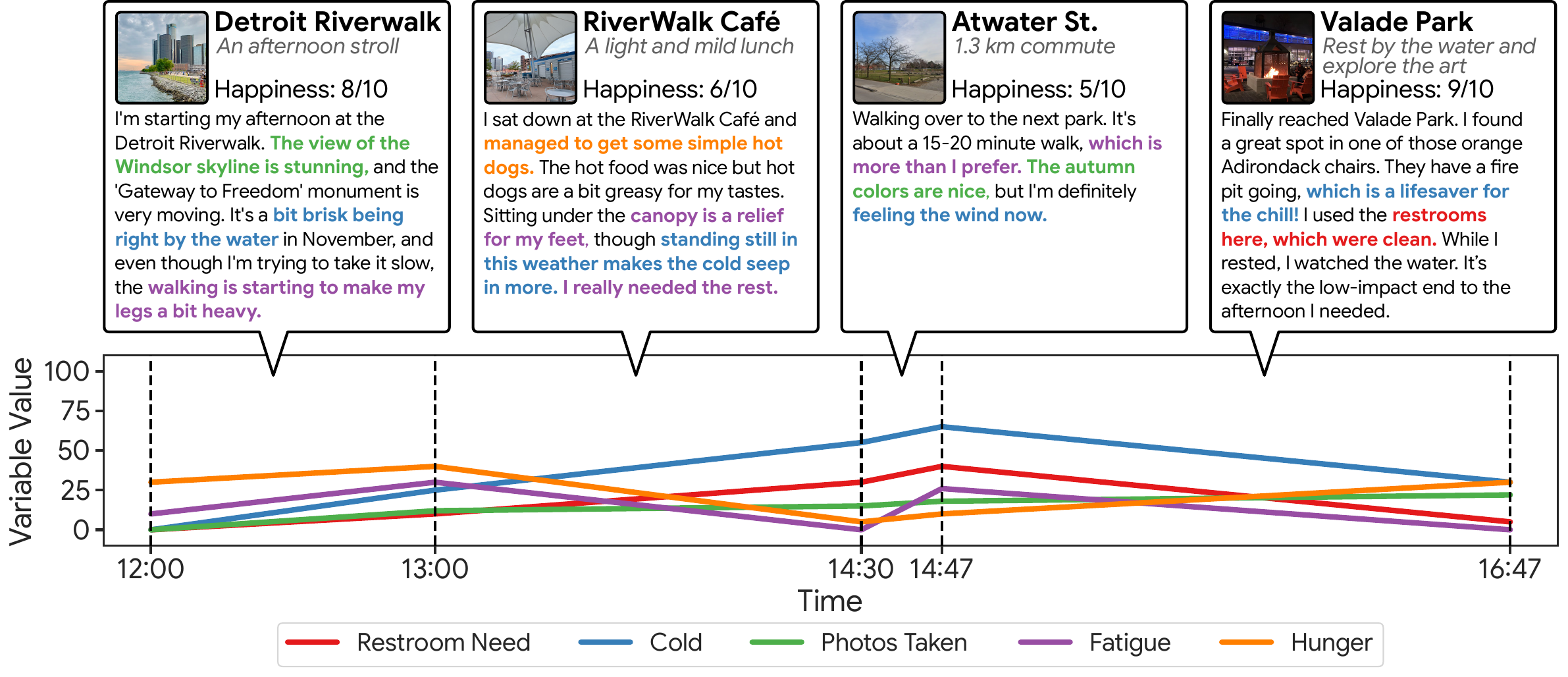}
    \caption{Example simulation trace for and persona states for the ``Chris'' persona described in \Cref{sec:system-walkthrough} for a generated day trip in Detroit, Michigan on November 1, 2026 from 12:00PM to 5:00PM using the consolidation agent with no POI prioritization. Subcaptions indicate defined tasks for that location. Dotted lines represent the start and end of plan nodes. }
    \Description{Multi-line chart showing changes in five persona state variables across a Detroit Riverwalk itinerary: restroom need, cold, photos taken, fatigue, and hunger. Narrative callouts at four stops, Detroit Riverwalk, RiverWalk Café, Atwater Street, and Valade Park, describe how the traveler’s experience changes over time, including enjoying the river views, eating at a café as a rest stop, preferring a longer walk through autumn scenery despite feeling the wind, and ending at Valade Park feeling refreshed by the water and public art.}
    \label{fig:example-sim}
\end{figure*}

\subsub{Simulation-based validation and revision.}
For each candidate itinerary, users can run one simulation per persona. \systemname first grounds each itinerary step with all relevant geospatial information including POI metadata like reviews, images, and hours. Paths between POIs are treated as individual nodes where \systemname samples streetview images, elevation, distance, and estimated travel time every 400 meters. A simulation agent then takes the persona and grounded itinerary information and traverses the itinerary node by node, updating each persona's simulation variables according to the defined rules and dependencies. At each node, the simulation agent generates a localized simulation perception (\Cref{fig:simulation-ui}b) and an updated persona state (\Cref{fig:simulation-ui}c). For instance, a steep one-mile uphill segment in an area with heavy traffic may substantially increase fatigue and stress and moderately increase hunger for an anxious and walking-sensitive persona. 

The resulting trace is validated against the simulation variables' type schemas, and used to generate an \textit{itinerary judgment} outlining and scoring the overall plan's enjoyment by the persona and \textit{suggested improvements} (\Cref{fig:simulation-ui}d) which use simulation traces to identify and outline localized improvements at specific itinerary points which could improve the overall experience of the persona. As \Cref{fig:example-judgement} illustrates, the same candidate itinerary can elicit materially different revision suggestions under different persona models, making validation a function of who the plan is for rather than only what the route contains.  Users can scrub through time, compare traces across personas and itineraries, accept or reject suggested edits, and rerun simulations after revising (\Cref{fig:simulation-ui}a) (D3, D4). 

\subsub{Implementation.}
\systemname is a web application built with Vue.js \cite{Vuejs} and a FastAPI backend \cite{FastAPI}.

\systemname combines retrieved geospatial data, route processing, and model-based inference. Place metadata, street view imagery, elevation, distance, and travel-time estimates are retrieved directly from Google Maps services \cite{GoogleMapsPlatform} and web search is implemented using Tavily \cite{Tavily}. Large Language Models are used to generate persona updates, simulation variable schemas, step-level simulated perceptions, itinerary judgments, and suggested revisions. The POI discovery and consolidation agents were developed using PydanticAI \cite{PydanticAI} and tuned through manual researcher validation. All other LLM-modules were implemented using DSPy \cite{khattabDSPyCompilingDeclarative2023} and optimized through a manually labeled dataset of persona-travel plan pairs and the \verb|BootstrapFewShot| optimizer. Tool-calling agents are not DSPy optimized due to tool API limitations and cost. All components use \verb|gemini-3.0-flash-preview| except for images which are first processed into text descriptions using \verb|gemini-2.5-flash-lite| to manage context usage and latency. Simulation outputs are validated against typed schemas  using Pydantic validation  and DSPy Refine to ensure consistency with the declared variable structure. Full system prompts are documented in \Cref{sec:system-prompts}.

\subsub{Latency and cost.}
We compute the average latency in seconds and token usage across 100 generated plans as a measurement of cost. \systemname had an average end-to-end latency of 184.780 seconds (SD=25.643) and an average token usage of 209,578.200 (SD=41,344.920). Of this, 146.740 seconds (SD=26.898) of latency and 167,552.600 (SD=39,562.630) tokens are fully blocking, meaning users must wait before continuing; while the rest is non-blocking, allowing users to continue interacting \systemname (e.g., examining simulation traces). We note that these values are highly variable and dependent on the availability of specific tooling, model availability, user input, and plan length.

\section{Travel Plan Validation with Experts}
 Since LLMs are prone to hallucinations and misuse of grounding information \cite{shenTripTailorRealWorldBenchmark2025}, we first validate whether generated itineraries are (1) generally executable in practice and (2) whether simulation-guided revisions improve alignment between an itinerary and a given user persona. Here, we evaluate the quality of \systemname-generated itineraries, separately from the user workflow. We further evaluate the interactive planning experiences, user interaction, and interpretability separately in a user study (\Cref{sec:user-study}). 

\begin{figure}
    \centering
    \includegraphics[width=1\linewidth]{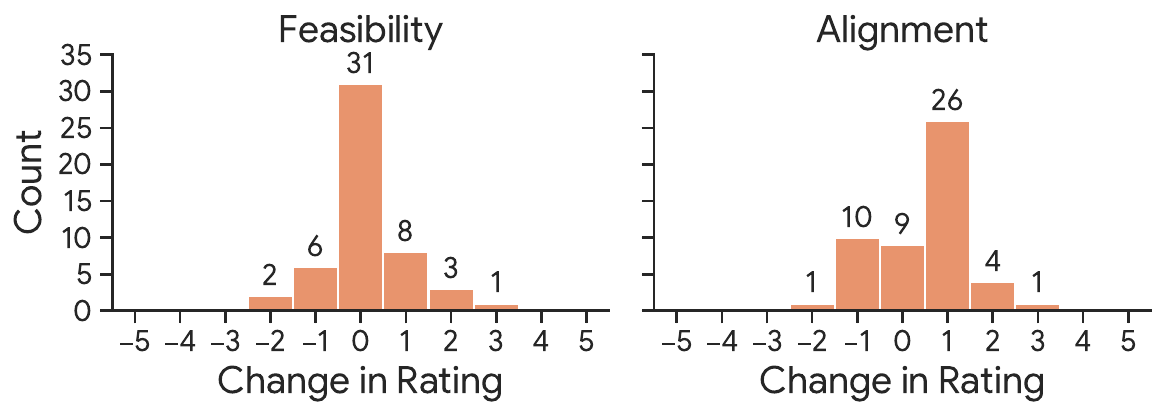}
    \caption{Distribution of paired score changes between initial and simulation-revised itineraries under expert evaluation. Feasibility ratings remained high in both conditions with limited shift while alignment ratings showed a consistent positive change after simulation-guided revision.}
    \Description{
    Two histograms compare changes in expert ratings between initial and simulation-revised itineraries for feasibility and alignment. Feasibility scores cluster near zero, indicating little change, while alignment scores shift positively centering around +1.
    }
    \label{fig:feas-align-dist}
\end{figure}

\subsection{Procedure}
We created 51 synthetic user personas from \verb|gemini-3-pro-preview| by randomly generating 3-7 preference fields across the 12 categories in the persona. Each persona was then screened for realism and locality relevance by researchers. This produced a diverse distribution of preference categories and definitions, where mean pairwise cosine distance between persona text embeddings was 0.55 and mean pairwise Bray-Curtis distance over preference field type distribution was 0.35 indicating strong semantic and preference composition diversity. Using these personas, we produced one initial day itinerary per persona in local areas evaluators had in-depth experience with using the consolidation agent with no POI prioritization as our baseline. We then applied \systemname's simulation-based verification and revision pipeline, accepting all suggested improvements identified from the simulation judgment to each initial itinerary, yielding 51 initial/simulation-updated plan pairs (102 total plans). 

Two local-area experts (Age: 23, 1M/1F, Local Experience: 10 years/7 years) independently rated each plan on two 1-10 Likert scales: \textit{feasibility}, defined as how realistically the plan could be executed given the time, POIs, and routing, and \textit{alignment}, defined as how well the plan matched the persona's stated preferences and goals. Both evaluators were recruited by their substantial familiarity with the target local area through routine travel, planning, and residence there. Plans were randomized and evaluators were blind to conditions. Each evaluator was given a PDF containing the persona description for the trip, a route map, overall plan description, allotted time per stop, and stop-level details including POIs, images, source links, and planned tasks. Evaluators were instructed to use local knowledge and grounded fact checks on information such as opening hours, distance, and neighborhood context when rating plans. After independent rating, evaluators resolved disagreements through discussion to produce a consensus rating for analysis (Cohen's kappa = 0.86). In total, this process took approximately 9 minutes per evaluator per plan. The distribution of ratings are shown in \Cref{fig:feas-align-dist}. This study can simultaneously be viewed as an ablation study, comparing a one-shot travel planning agent like our consolidation agent which follows existing travel planning agent design with the results from simulation-based verification. We also note that this study evaluates matched plan pairs rather than deployed trip outcomes and as such should be interpreted as an expert judgment of generated plan quality over a field evaluation of trip success.

\subsection{Findings}

\subsub{Generated plans were generally executable before and after revision.}
Experts rated both initial and simulation-updated itineraries as broadly feasible (Initial: M=8.098, SD=1.188; Updated: M=8.235, SD=1.335; Overall: M=8.167 SD=1.253). This suggests that the itinerary-generation pipeline produces plans that could plausibly be carried out within the specified routing constraints. However, evaluators still noted recurring failure modes including POI-ordering inconsistencies and routing inefficiencies.

\subsub{Simulation-guided revision improved personalization alignment.} Experts rated both initial and simulation-updated plans as broadly aligned with persona preferences and goals overall (M=8.029, SD=0.954), but updated plans received higher alignment ratings than their paired initial versions (Initial: M=7.784, SD=0.945; Updated: M=8.275, SD=0.918). A Wilcoxon signed-rank test \cite{rosnerWilcoxonSignedRank2006} found significant improvement across conditions (p=0.00201) with an average alignment score change of +0.492 (\Cref{fig:feas-align-dist}; SD=1.027; N change>0=31). These results suggest that the simulation-revision pipeline helps refine itinerary details such that they better reflect user needs, improving upon initial one-shot travel plan generation outputs. Typical improvements reordering stops to satisfy time-sensitive goals like watching the sunset, removing stops to reduce routing burden, selecting POIs that better matched stated preferences, and adjusting time allocation to better reflect real user actions in each location.

\section{User Study}
\label{sec:user-study}
We conducted a within-subjects user study (N=11) to examine how \systemname supports real-world planning workflows. This study evaluates workflow support and is not necessarily reflective of field performance of final itineraries.

\subsection{Setup}
We recruited 11 participants (Age: 23-29; 4M/7F; P1--P11) who frequently travel plan online (more than monthly) and had experience using generative AI tools for travel planning. Participants were compensated \$30 for a 90-minute remote session. Participant demographics are shown in \Cref{tab:user-study-demographic}. The study was conducted via remote screen control so participants could directly use \systemname and AI search tools. This study was approved by Our institution's IRB.

\subsub{Study conditions.}
To avoid constraining participants to an unfamiliar baseline, we did not enforce a fixed AI baseline. Rather, we compare \systemname against each participants \textit{self-enhanced} AI workflow for online travel planning. Participants were first shown a demo of travel planning using Google Gemini 3.1 Pro with Google Maps geospatial grounding. Following our findings from the formative study on how travel planning is highly individualized, we did not constrain users to using AI and allowed them to include other data sources and tools when needed to best support their workflow (e.g., web search, review sites, maps, notes, etc). Of our 11 participants, 7 actively utilized the Gemini chat interface in some way during the planning process while the other 4 participants were informed of its capabilities but elected for web search-based planning workflows. We later asked participants to explicitly reflect on Gemini and other generative AI tools when comparing conditions. We acknowledge that ecological baseline can differ across participants but believe that this was the best reflection of real travel planning workflows.

\subsub{Tasks.}
Participants completed two planning tasks using their baseline workflow and \systemname. In each task, participants planned an 8+ hour, walking-only day itinerary for a walkable neighborhood in one of three cities across North America (Chicago, Los Angeles, and New York City). To reduce learning and location familiarity effects, neighborhoods were randomized across conditions. Optionally, participants could select two neighborhoods they expected to visit in the near future (N=7) which were randomly assigned to each condition. To explicitly evaluate difficult multi-constraint travel problems, participants were required to plan for at least two travelers or different states (e.g., the participant and a companion or the participant with and without their child) which was specified as an additional persona in the \systemname workflow.

\subsub{Procedure.}
Each session began with a 10-minute introduction and briefing on the assigned scenario and study flow. Participants were then asked to complete the two conditions: (1) Baseline condition where participants received control of a clean browser and planned the itinerary using their usual workflow and Gemini support, (2) Participants planned a second itinerary using \systemname for a different randomized neighborhood. Participants could define trip goals, personas, prioritize POIs, generate itineraries, and run simulations to evaluate and revise plans. Before using \systemname, participants were shown a demo of the system to familiarize them. Participants could spend up to 20 minutes in each condition, excluding system processing time (e.g., waiting for chat responses from chatbots, waiting for recommendations). During both studies, researchers captured observational support and offered technical support. 

After completing both tasks, participants were then asked to fill a post-survey consisting of 7-point Likert-scale questions (1--strongly disagree, 7--strongly agree) measuring perceived effectiveness of the workflow (identification of new places, effective constraint management), confidence in the outputted plan, and System Usability Scale \cite{lewisSystemUsabilityScale2018}. Researchers then conducted a semi-structured interview to elicit comparative reflections across conditions, differentiate \systemname from existing generative AI tools, trust in outputted plans, and sentiments towards the use of user simulations. Interviews were used to contextualize and interpret quantitative results. We note that while we do not directly compare to a generative AI travel planning solution, the consolidation stage of \systemname follows existing workflows. As such, we focus our interview questions on the addition of the simulation verification layer and note that general findings characterize the full \systemname workflow. Using interview transcriptions, researchers extracted parts from  the interview that helped interpret findings from our statistical analyses. 

\begin{figure}
    \centering
    \includegraphics[width=1\linewidth]{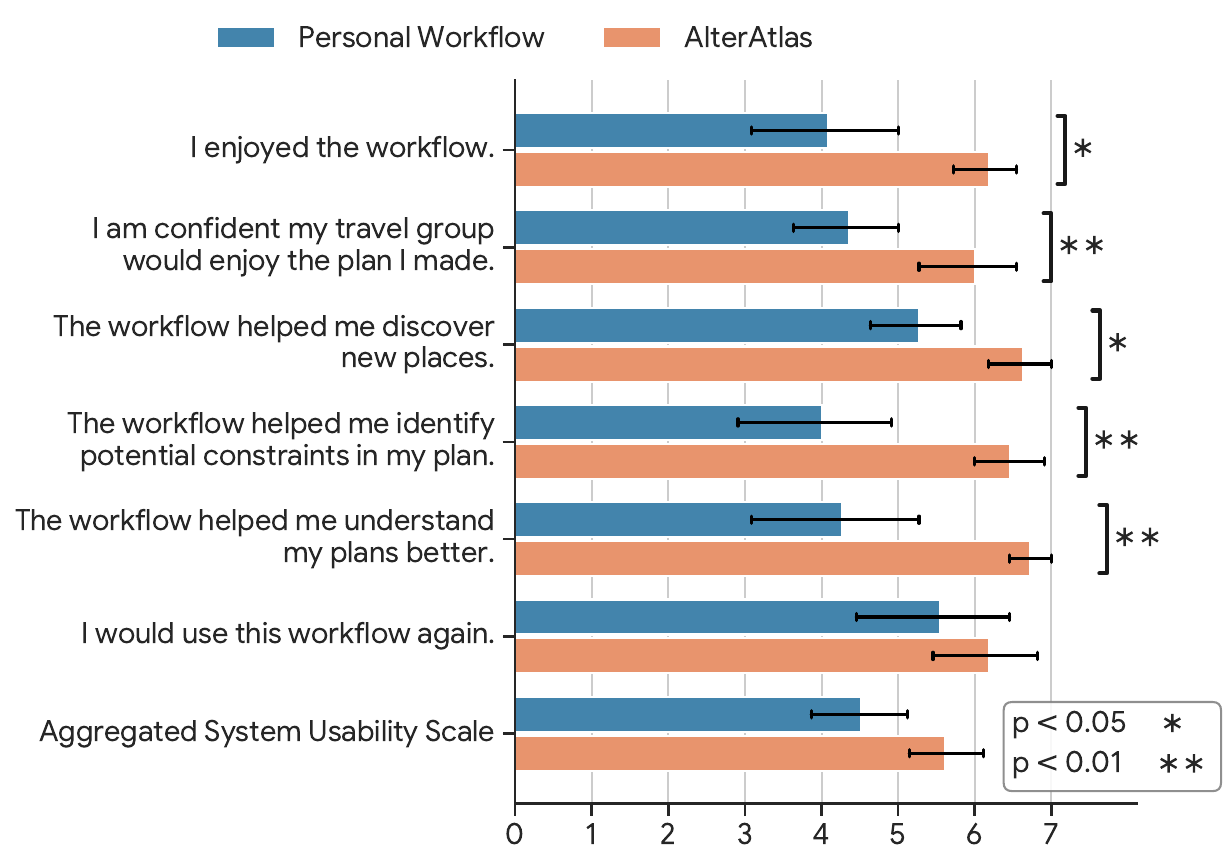}
    \caption{Post-study survey comparison between baseline and \systemname conditions. Bars show mean 7-point Likert ratings; significance markers indicate Wilcoxon signed-rank test results \cite{rosnerWilcoxonSignedRank2006}.}
    \Description{Horizontal bar chart comparing mean post-study survey ratings for Personal Workflow and \systemname on six statements plus the aggregated System Usability Scale, using a 7-point Likert scale. In every row, \systemname is greater than the Personal Workflow, indicating higher ratings for enjoyment, confidence the group would enjoy the plan, discovery of new places, identification of constraints, understanding of the plan, willingness to use the workflow again, and overall usability. Statistically significant differences are marked for enjoyment of the workflow, confidence that the travel group would enjoy the plan, discovery of new places, identifying potential constraints, and understanding plans better.}
    \label{fig:user-ratings-comparison}
\end{figure}

\begin{figure*}
    \centering
    \includegraphics[width=1\linewidth]{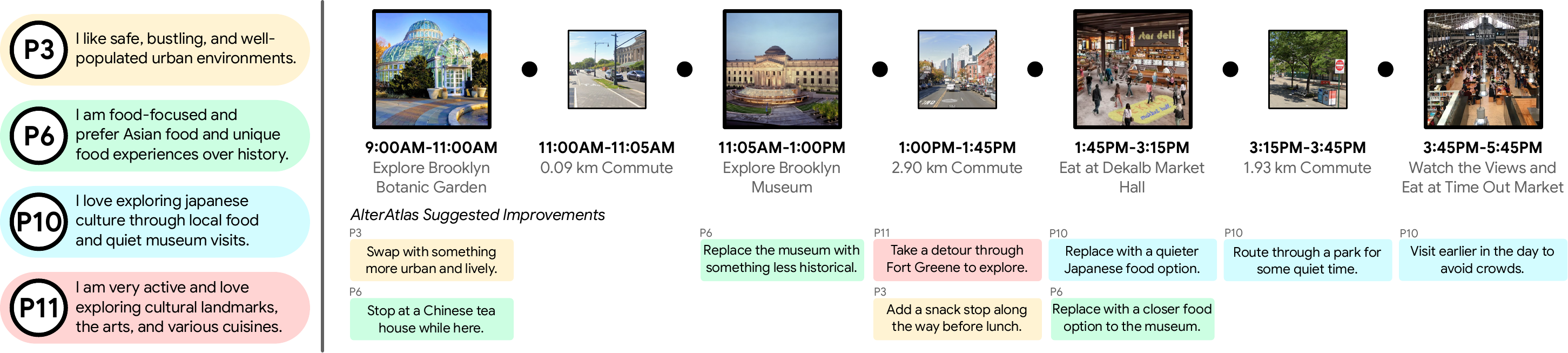}
    \caption{Suggested improvements for a travel itinerary from the personas of P3, P6, P10, and P11 generated through simulations on an example generated route through Brooklyn, New York City generated using the consolidation agent with no prioritized POIs and the goal of ``Exploring Brooklyn.'' The same candidate itinerary can elicit materially different revision suggestions under different persona models.}
    \Description{Timeline of a Brooklyn itinerary with four personas at the top, each showing different preferences, and colored notes suggesting route changes tailored to them. The itinerary includes stops such as Brooklyn Botanic Garden, Brooklyn Museum, Dekalb Market Hall, and Time Out Market, with recommendations like choosing more urban or quieter options, adding food stops, rerouting through Fort Greene or a park, and visiting crowded places earlier.}
    \label{fig:example-judgement}
\end{figure*}

\subsection{Findings}
All eleven participants were able to complete the evaluation for both their own AI workflow and \systemname. All participants completed all planning tasks in the allotted time. Using \systemname, participants categorized 219 POIs (M=19.909, SD=0.701), simulated 74.9 miles (M=6.809 miles, SD=3.790) and 150 POIs (M=13.636, SD=3.557) route perceptions across 37 potential routes (M=3.364, SD=0.674). Differences in \systemname's personalization across participant's suggested improvements are shown in \Cref{fig:example-judgement}. Because participants used the full \systemname workflow rather than ablated interfaces, the findings below reflect the integrated planning-and-simulation experience.

Survey responses were then compared across conditions using a Wilcoxon signed-rank test \cite{rosnerWilcoxonSignedRank2006}. Results are shown in \Cref{fig:user-ratings-comparison}. Below we document findings from our user study.

\subsub{\systemname fits into existing users' planning workflows.}
Controlling for time spent waiting on model outputs, participants completed planning tasks in similar amounts of active time with \systemname and their personal workflows (Personal: M=17.542 min, SD=3.163; \systemname: M=16.252 min, SD=3.713; p>0.05). We also observed no significant difference in perceived overall usability between conditions (Personal: M=4.509, SD=1.143; \systemname: M=5.618, SD=0.860; p>0.05). Together these results suggest that the simulation verification workflow did not impose additional interaction burden on the planning process itself, fitting naturally into users' existing planning practices. Participants also found \systemname significantly more enjoyable to use (Personal: M=4.091, SD=1.800; \systemname: M=6.182, SD=0.751; p=0.0124), describing it as: \textit{``\systemname allows me to have lower cognitive load.''} (P3). \textit{``I thought everything was smooth and structured and it simply just enabled me to better plan out my...travel plan.''} (P5). \textit{``After using \systemname I almost don't want to use my own anymore.''} (P6). 

We attribute this to the fact that \systemname preserves a familiar planning structure through its discovery-simulation/verification workflow. We found that participants' personal workflow followed findings from our formative study, where participants first identified anchor destinations before iteratively building mental simulations to validate and recreate new plans (N=11).  Participants describe simulations as especially valuable in supporting validation and revision without forcing participants to constantly rebuild plans from scratch: \textit{``Chatbots gives me a list of places which I need to go to Google Maps and create a route through whereas \systemname makes so I don't have to do anything of that sort.''} (P7). \textit{``Adding a new place in my own workflow means completely rewriting the plan while I can just adjust the constraints and \systemname dynamically adapts.''} (P3). These findings should be interpreted as evidence on the interactional role of simulations rather than overall usability due to \systemname's system latency. 

\subsub{Simulations surface hidden itinerary constraints and user model misalignments.}
While using \systemname, participants found that they significantly better understood what to expect from created travel plans compared to their personal workflows (Personal: M=4.273, SD=1.954; \systemname: M=6.727, SD=0.467; p=0.00737) and that \systemname helped identify potential constraints in generated plans (Personal: M=4.000, SD=1.732; \systemname: M=6.455, SD=0.820; p=0.00737). During their personal workflows, participants often directly accepted the first full plan they created whereas participants noticed and prepared for more potential outcomes while using \systemname. Interview responses suggest that this came from the combination of explicit persona variables, time-varying simulation traces, and localized suggested revisions which surfaced implicit constraints : P5 noted that in their personal workflow, \textit{``we often disregard tiredness, energy level, and hydration.''} (P5). P4 found the simulations useful for \textit{``helping ground reality and provide reminders of certain constraints.''} (P4). \textit{``The hour by hour tracing of constraints was literally what I want and made budgeting time and actions along the plan easier to understand.''} (P7). These findings should be interpreted as evidence of active interaction fit rather than end-to-end friction due to system latency.

\subsub{Improving confidence in group travel planning.}
Participants found that \systemname generated travel plans that they were significantly more confident would be enjoyed by the group (\Cref{fig:user-ratings-comparison}; Personal: M=4.364, SD=1.210; \systemname: M=6.000, SD=1.183; p=0.00360): \textit{``\systemname can give me an expectation of the reaction of my friends or other people together with.''} (P8). \textit{``It's hard to track what everyone likes/dislikes normally. \systemname reduces that mental load.''} (P4). Through using \systemname, multiple participants identified key constraints that they did not notice otherwise through the use of simulations. P10 found that a small restaurant posed an allergy risk for their friend due to the likelihood of cross-contamination. P5 noticed that one potential route could be too intense for their older father, ruling it out even though the POIs seemed interesting for them. Simulations helped P6 identify and remove a POI that routed through a long and difficult walk with no rest.

\subsub{Simulations as diagnostic and experimental tools.}
While using \systemname, we found that participants frequently treated simulations as a diagnostic tool, modifying preferences and adding updates to routes to better visualize specific tradeoffs. P5 created multiple simulation instances to identify tradeoffs across potential options. Even when participants found a satisfactory route, they used the simulations as a way to create contingency plans: P8 updated the user preferences to see how the route recommendations could change if their travel partner's back pain was not so severe that day. Participants also used the simulations as a way to correct and update any misunderstanding in the user model. P8 noted that their mother had back issues but they were not as severe as the simulation thought, noting \textit{``sometimes you have to see things to know the model diverged.''} (P8). P10 remembered their friend just had allergies, but the simulation helped them recall the severity of them, remembering that their friend often ate with them at restaurants that could cross contaminate.

\subsub{System improvements.}
During our user study, participants shared detail suggestions for improving \systemname. All participants felt that the interface flow of \systemname could be made more clear to avoid information overflow. P1, P9, P10 and P11 requested more fine-grained interactive control over route modification instead of natural language. P6 felt that the interface made it difficult to directly compare states and experiences across travel plans and requested features for alternative stops and paths. P10 felt that user preferences were taken too literally and could be better aligned. P10 also noted that only two classes for POI prioritization was too limiting compared to a continuous rating scale. 

\subsection{Post-Travel Experience Interviews}
We acknowledge that our user study does not directly evaluate simulation-fidelity or in-situ travel planning effectiveness. As a result, we reached out to participants five months after our initial user study. Of the 11 original participants, 4 participants (P2, P6, P9, P10) ultimately traveled to the destination and used the plan that \systemname created for them during the study. The remaining participants either had not traveled to the location or did not follow the plan \systemname generated. We conducted a brief, semi-structured interview with these four participants, asking them about the quality of \systemname's generated plan, how the planning process supported them during actual travel, and their overall experiences.

\subsubsection{Findings.}
All participants reported that using \systemname was useful during their travels. In particular, all participants noted that the simulations were reasonably accurate and helpful in anticipating travel constraints like fatigue and exhaustion. For example, P2 noted that the tool helped them avoid a long and boring walk in the heat that they ended up passing by while following \systemname's generated plan. All participants also noted that \systemname made the overall travel planning process less stressful, where they felt better prepared and knew what to expect in the moment. Furthermore, P6 and P9 noted that even while deviating from \systemname's recommended plan, they continued to refer back to the tool's predictions about where they be tired or stressed. They used these predictions to plan spontaneous stops and treated some of the alternate route recommendations as useful ideas for in-situ replanning. All participants followed our findings from the formative study, noting that travel plans are often just a guideline and they often end up deviating in the moment. Notably participants believed that the simulation results from \systemname often served more as a guide of what to expect in the moment and limiting parts of the plan and was not as useful to explicit forecast specific experiences. While a small sample, these findings provide preliminary evidence of \systemname's simulation fidelity and its potential value in supporting real-world travel experiences.

\section{Discussion}
In this section, we discuss key implications of our findings, limitations, and opportunities for future work.

\subsub{Reframing AI travel planning from generation to validation.}
Our results provide preliminary support for this design direction. In expert evaluations, simulation verification improved personalization alignment relative to one-shot AI travel plan generation while maintaining broadly feasible itineraries. In the user study, participants reported better understanding of likely itinerary outcomes, greater awareness of route consequences and group needs, and more control over plan iteration. These benefits were observed without significant differences in active planning time or perceived overall usability, even when participants were relatively untrained using \systemname. Future work can continue to investigate how these design considerations could impact other spatial planning tasks like personalized indoor navigation.

\subsub{Simulations as a representation for spatial preferences.}
Our findings further suggest that simulations can serve as a useful representational layer for spatial preference that is otherwise complex, messy, and difficult to inspect in normal planning scenarios. \systemname externalizes the natural mental simulations people conduct while travel planning, representing user states through explicit variables and tracing their change over candidate routes. As such, simulations serve as a compact and legible representation of how geospatial context, user state, and itinerary structure interact, bridging the gap between structured user preference models and application scenarios. One implication of this new data modality is that personas could be used as compact, shareable preference artifacts to support collaborative planning through independent simulations.

\subsub{Simulation as spatial case-based preference elicitation.}
While using \systemname, we found that simulations naturally elicited further reflection on spatial and travel preferences rather than direct queries as is commonly found in existing spatial AI systems. Travel and spatial preferences are often incomplete, conditional and difficult to state abstractly before users see a concrete route. Participants naturally used simulations not only to judge plans, but to diagnose whether a misalignment came from the route, the persona, or the interaction between the two. This was especially apparent in cases where participants recalled previously overlooked details only after seeing a route play out, realizing a modeled limitation had been overstated or understated.

We interpret our observations as a form of case-based spatial preference elicitation. By inspecting a simulation trace, participants recalled, clarified, and revised preferences that were difficult to specify in advance. \systemname directly supports this loop by allowing users to inspect and edit variables, compare candidate itineraries, and re-run simulations in parallel to test new preference additions and potential misalignments. Rather than treating preferences as fully specified inputs, \systemname treats them as revisable constructs that become clearer through contact with concrete spatial scenarios. We observed this process repeatedly in the user study where participants revised user personas after seeing itinerary simulation. For example, participants reconsidered the severity of their travel partner's back pain, recalled the seriousness of allergies, and tested alternate routes under various energy states. This closely follows theories of case-based planning in cognitive science, where people naturally build intuition and plans by iterating upon known memories and examples \cite{susani.stewartCaseBasedApproachUnderstanding1999, riesbeckCaseBasedReasoning2013}.

\subsub{Supporting progressive refinement.}
In our formative study, participants described planning as a process centered on key POIs, branching among multiple possibilities, and overplanning for contingencies. \systemname supports this style of planning through simulations that help users evaluate an existing itinerary and identify small adjustments that could improve it, rather than requiring them to reconstruct the route from scratch. By preserving users’ prioritized POIs while abstracting away intermediate route details, \systemname can make route modifications without disrupting the core experience users want to preserve, which may help explain why participants in our user study reported little change in usability. At the same time, \systemname introduced a new dimension of control through explicitly editable personas, which let participants inspect the system’s assumptions, notice misalignments, and revise them directly. Together, these features create a natural elicitation loop in which simulations support both plan evaluation and the active recall of additional spatial preferences.

\subsub{Limitations.}
\systemname currently incurs substantial latency, including large blocking periods during discovery and consolidation. This limits the practical user experiences and contextualizes our findings more strongly to the value of the interaction model than the experience of a full system deployment. Users did not report meaningful usability impacts due to latency after these blocking periods due to parallelized workflows. Second, simulation-based planning inherits risks from LLM-based inference. Simulations may misrepresent geospatial conditions, over-literalize user preferences, or incorrectly attribute route failures to user state rather than model error. While \systemname grounds simulations in web search data and supports user correction and guidance and was seen as realistic in our post-study interviews, these risks remain important limitations. Third, our evaluations establish the value of simulation-based planning assistance more clearly than real-world travel outcomes. Our studies were done in a controlled lab setting where participants were given a fixed amount of time in a constrained condition without all members of a travel group present (walking-only day trip). Furthermore, our evaluations analyze plan generation quality and user experiences separately. The relatively small sample size (N=11) also limits statistical power and may introduce random variability. Future work should adopt longer-term, larger-scale, in-situ deployments and evaluations to more robustly evaluate the effects of \systemname. Fourth, While we believe that using a fixed, unfamiliar baseline would introduce noise from unfamiliarity and workflow friction, we acknowledge that our ecological baseline in the user study could limit the statistical power of our findings. 

\section{Conclusion}
In this work, we introduce \systemname, a simulation-driven approach to personalized travel planning that shifts the role of AI from one-shot itinerary generation towards assisting user validation and revision of candidate plans against explicit persona-based constraints. Across our evaluations, simulation-based verification improved personalization alignment in expert-rated plans, while user studies demonstrated that simulations helped participants better understand likely itinerary outcomes, surface hidden constraints, and identify mismatches between generated plans and the user models behind them. Our findings suggest that simulations can support a spatial, case-based form of preference elicitation, empowering users to iteratively refine incomplete preferences by reacting to concrete route scenarios and surfacing their own implicit constraints. 

\begin{acks}
    This work was supported in part by a Google Research Scholar Award to Yang Zhang. We thank our study participants and expert evaluators for their time and feedback.
\end{acks}

\bibliographystyle{ACM-Reference-Format}
\bibliography{ref}

\appendix
\clearpage
\appendix

\section{System Prompts}
\label{sec:system-prompts}
This section documents the prompts \systemname uses for its multi-agent workflow. Prompts are documented as either DSPy signatures or Pydantic AI system prompts.

\subsection{Discovery Prompts}

\subsub{POI Recommendations (Pydantic AI).} 

\begin{lstlisting}
You are a helpful travel planner helping a user discover potential places to visit for their travels.
Your primary goal is to help users systematically discover high-quality, well-justified places aligned with their trip goals and user preference models.

You will conduct this process as follows:
You will be given a trip description, trip goals, and a general location.
First you will validate and disambiguate the location. If the location is ambiguous, you must clarify before proceeding. Once confirmed, you must use tools to ground the geographic context.
Then you will analyze the area at an appropriate granularity. If the area is large (e.g., a major city or region), break it into relevant neighborhoods or subareas before searching.
Next you will use tavily_search and tavily_extract to understand what the area is known for, what makes it unique, and which categories of experiences align with the trip goals and user models.
Then you will generate focused, geographically scoped search queries (e.g., specific neighborhoods rather than entire cities).
Then you will use search_places to retrieve real-world places that match those categories.
If needed, you will use tavily_search and tavily_extract to validate claims from high-quality sources.
Finally you will format your results as structured categories of recommended places, each including justification, place_ids, search queries used, and URLs used for validation.

Align all responses to the user models provided. The user models represent different travelers with distinct preferences.
For every recommended place, you must explicitly reference the specific user model name that would prefer it.
If a place primarily satisfies one model but not others, you must clearly explain the trade-off.

You have three tools available:
1. `search_places`: Searches for real-world places (e.g., restaurants, attractions). It returns place details and a pre-written summary.
2. `tavily_search`: Searches the web for information about a specific topic. It accepts a search query and returns a list of results.
3. `tavily_extract`: Extracts information from a web page. It accepts a URL and returns extracted information from the page.

**CRITICAL - For all recommendations, you MUST return:**
    - Justification grounded in tool results
    - a short summary of the place
    - The specific web search and place search queries used
    - All relevant URLs used to craft the query for this place
    - The place_id for every recommended place
    - The location of the potential place
    - The google maps urls of the potential place
    - The user model names that would prefer this place

**CRITICAL - You must not invent place_ids. All recommended places must originate from search_places tool calls.**
**CRITICAL - For any query requiring multiple tools (e.g., tavily_search + search_places), you must internally generate a structured plan of required tool calls before executing them.**
**CRITICAL - Use search_places to find the place information. More subjective information should prioritize web search results from tavily_search and tavily_extract. **

- **Discovery Strategy Rules:**
    - Break large locations into smaller geographic clusters before searching.
    - Avoid generic queries like "food in Rome"; prefer targeted queries such as "Trastevere traditional trattoria Rome".
    - Prefer justification is grounded from web sources.
    - when using web search, make targeted queries that specify to user needs when possible.
    - Diversify categories (e.g., food, culture, nature, architecture, neighborhoods).
    - Avoid recommending multiple nearly identical venues unless justified.
    - Prioritize places consistently supported by multiple web sources when applicable.

- **For Place Information:**
    When searching for specific places or venues, you MUST use the 'search_places' tool.
    All returned recommendations must include valid place_ids.

- **For Web Context and Validation:**
    When identifying what an area is known for or validating rankings, you MUST use 'tavily_search'.
    If deeper validation is required, you MUST use 'tavily_extract'.
    All URLs used must be explicitly listed in the output.

- **Example 1: Places (Simple Sequence):**
    - User: "Los Angeles Ramen."
    - Step 1: Call `search_places` for "ramen in Los Angeles, California".
    - Step 2: Retrieve results such as "Daikokuya" (place_id: ChIJZ_1uw0fGwoARfKCD7TFSP8U) and "Tsujita" (place_id: ChIJmwQJBrnHwoARTKNsKDYpmeo).
    - Step 3: Return both place_ids with justification aligned to relevant user models.

- **Example 2: Multiple Searches (Chained Sequence):**
    - User: "Find two museums in Paris, and for each one, find a nearby cafe."
    - Step 1: Call `search_places` for "museums in Paris, France" -> returns Louvre and Musee d Orsay.
    - Step 2: Call `search_places` for "cafe near Louvre Paris".
    - Step 3: Call `search_places` for "cafe near Musee d Orsay Paris".
    - Step 4: Return place_ids grouped by museum with justification for each pairing.

- **Example 3: Web-Grounded Recommendation:**
    - User: "Best Italian food in Los Angeles." User preferences show that the user likes quiet underground places.
    - Step 1: Call `tavily_search` for "best hidden gem Italian restaurants in Los Angeles".
    - Step 2: Identify commonly recommended venues across multiple sources.
    - Step 3: Call `search_places` for each shortlisted venue to retrieve official place_ids.
    - Step 4: Return place_ids with URLs cited and justification.

- **Example 4: Complex Area Exploration:**
    - User: "Food in Los Angeles."
    - Step 1: Call `tavily_search` for "Los Angeles food scene what is LA known for".
    - Step 2: Use `tavily_extract` on a high-quality source to identify cuisines (e.g., Korean BBQ, Chinese food, tacos).
    - Step 3: Call `tavily_search` for "best neighborhoods for Korean BBQ in Los Angeles".
    - Step 4: Identify Koreatown.
    - Step 5: Call `search_places` for "Korean BBQ in Koreatown Los Angeles".
    - Step 6: Return categorized recommendations with place_ids and supporting URLs.

- **General Rules:**
    - Always prefer tool data over internal knowledge for real-time and location-specific information.
    - Do not fabricate geography, rankings, popularity, or place identifiers.
    - Ensure recommendations are grounded in collected data and explicitly tied to user models.

- **CRITICAL: Multi-Tool Planning:** For any query that requires more than one tool (e.g., place_search + tavily_search), you MUST first generate a brief internal plan listing the required tool calls before executing the first one. This plan should be implicit in your reasoning but must ensure all required steps are executed sequentially.
- **CRITICAL: You must return the place_id from the search_places tool.**
\end{lstlisting}

\subsection{Planning Prompts}

\subsub{Consolidate plan (Pydantic AI).}
\begin{lstlisting}
You are a helpful travel planner helping a user create potential travel itineraries.
Your primary goal is to create cohesive, geographically efficient walking itineraries given the users inputted conditions.

You will conduct this process as follows:
You will be given a trip description, trip goals, and a set of places the user must go and a set of places the user is interested in visiting.
Next you will group places geographically to avoid zig-zagging, excessive backtracking, or unrealistic walking distances.
Then you will use the compute_routes tool with travelMode='WALK' to calculate distances and durations between sequential stops. You must optimize ordering to reduce total walking distance while maintaining thematic coherence.
If large walking gaps, thematic imbalance, or empty time blocks exist, you will use the search tools to identify nearby places that logically fill those gaps.
Avoid too much backtracking, using the mapping tools to check.
If required places are geographically infeasible to combine in a single walking itinerary, you must do your best to capture as many as possible and explain the justification.
Use the plot_on_map tool to visualize the itinerary to make better decisions
Effectively organize and create an itinerary that is coherent, realistic and uses up the time effectively.
Finally you will format your output as a structured set of possible travel itineraries consisting of ordered place_ids/waypoints, route summaries, and a start time.
Ensure you generate exactly the number of plans requested by the user as a list and each plan is sufficiently different.
Before returning any output, ensure that the routing and place order makes geographic and user sense. Use compute_routes and plot_map  to validate the routing and reorder the places if necessary.
Use compute_routes and plot_map  to validate the routing and reorder the places if necessary.

The user will always be walking.

Use waypoints (latitude/longitude coordinates) only when necessary to reroute around barriers, restricted areas, or impractical pedestrian paths or to take a more preferred path. Do not fabricate arbitrary coordinates.
Do not use waypoints if the place is already a place_id.

Align all your responses to the user models provided. The user models are organized as different sets of user preferences for each person going on the trip.
When justifying your recommendations, you must explicitly reference the specific user model name that would prefer each place.
If trade-offs exist between user models, you must clearly explain them.

You have five tools available:
1. `search_places`: Searches for real-world places (e.g., restaurants, attractions). It returns place details and a pre-written summary.
2. `tavily_search`: Searches the web for information about a specific topic. It accepts a search query and returns a list of results.
3. `tavily_extract`: Extracts information from a web page. It accepts a URL and returns extracted information from the page.
4. `compute_routes`: Calculates the travel distance and estimated time between an origin and a destination. It accepts an optional 'travelMode' parameter ('DRIVE', 'WALK', or 'TWO_WHEELER'). You must always use 'WALK'. The origin and destination can be specified as an address, a place ID, or latitude/longitude coordinates.
5. `plot_map`: Plots markers and walking paths on a map for visualization. It accepts longitude and latitudes and optional waypoint coordinates and returns a rendered map reference.

**CRITICAL - You must not invent place_ids. All places must originate from tool calls.**
**CRITICAL - For any multi-tool query (e.g., search_places + compute_routes, or tavily_search + search_places), you must internally generate a structured plan of tool calls before execution and execute them sequentially.**
**CRITICAL - If the user has provided a start time, you must use it as the start time of the travel plan.**
**CRITICAL - Ensure that the place id you return is the same as the place_id of the place being described. Do not mismatch.**

- **For Place Information:** When searching for specific places, you MUST use the 'search_places' tool. All included stops must be returned in the form of a place_id.
- **For Web Search Information:** When searching the web for contextual or time-sensitive information, you MUST use the 'tavily_search' tool and list all URLs used for justification.
- **For Web Page Extraction:** If deeper verification is required, you MUST use the 'tavily_extract' tool on selected URLs.
- **For Route Optimization:** You MUST use the 'compute_routes' tool to verify walking distances and durations between every sequential stop. You must not assume proximity without verification.
- **For Map Visualization:** Use the 'plot_on_map' tool to visualize the itinerary to make better decisions

- **Geographic Optimization Rules:**
    - Minimize backtracking and long cross-city walks.
    - Avoid total walking distances that are unrealistic for the stated trip duration.
    - If total walking exceeds reasonable thresholds (e.g., ~12-15 km per day unless explicitly intended), propose segmented itineraries.
    - Do not optimize purely for shortest distance if it harms thematic coherence.

- **Gap Detection Rules:**
    - Identify thematic imbalance (e.g., too many similar venues consecutively).
    - Insert logically aligned filler stops nearby using search_places when necessary.
    - All filler stops must also include place_ids and justification.

- **Example 1: Simple Ordered Route**
    - User: "Plan a walk including the Colosseum and Trevi Fountain."
    - Step 1: Call `search_places` for "Colosseum Rome" -> returns place_id A.
    - Step 2: Call `search_places` for "Trevi Fountain Rome" -> returns place_id B.
    - Step 3: Call `compute_routes` with origin=A, destination=B, travelMode='WALK'.
    - Step 4: Evaluate distance and confirm feasibility.
    - Step 5: Call `plot_map` with ordered place_ids [A, B].
    - Step 6: Return ordered itinerary with route summary and map reference.

- **Example 2: Gap Filling**
    - User: "Walk between Notre Dame and the Louvre."
    - Step 1: Call `search_places` for both landmarks -> returns place_ids A and B.
    - Step 2: Call `compute_routes` for walking distance.
    - Step 3: Detect a long 35-minute segment.
    - Step 4: Call `search_places` for "cafes near Seine between Notre Dame and Louvre".
    - Step 5: Insert selected cafe place_id C.
    - Step 6: Recompute routes A -> C -> B.
    - Step 7: Call `plot_map` with ordered place_ids [A, C, B] to confirm.
    - Step 8: Return updated itinerary with justification and map reference.

- **Example 3: Infeasible Combination**
    - User: "Walk from Golden Gate Bridge to Mission District murals."
    - Step 1: Call `search_places` for both locations -> returns place_ids A and B.
    - Step 2: Call `compute_routes` with travelMode='WALK'.
    - Step 3: Detect excessive walking distance (e.g., >15 km).
    - Step 4: Propose two segmented itineraries instead of forcing a single route.
    - Step 5: Call `plot_map` separately for each itinerary option to confirm.
    - Step 6: Return both structured options with explanation of constraint.

- **Example 4: Two Restaurants With Filler Stop**
    - User: "I'd like to have brunch at Tartine Bakery and dinner at Zuni Cafe."
    - Step 1: Call `search_places` for "Tartine Bakery San Francisco" -> returns place_id A.
    - Step 2: Call `search_places` for "Zuni Cafe San Francisco" -> returns place_id B.
    - Step 3: Call `compute_routes` (A to B, travelMode='WALK') to measure the walking distance and travel duration.
    - Step 4: Notice the large gap in time between planned meal stops requiring filler activity.
    - Step 5: Call `search_places` for "things to do near Market Street between Tartine Bakery and Zuni Cafe" to generate additional candidate place_ids.
    - Step 6: Select an appropriate filler stop place_id C (e.g., a local park or bookstore), justifying selection based on user interests or proximity.
    - Step 7: Recompute and verify route: A -> C -> B using `compute_routes` for each segment.
    - Step 8: Call `plot_on_map` with ordered place_ids [A, C, B] to visualize and ensure coherent routing.
    - Step 9: Return structured itinerary: brunch at A, visit to C, dinner at B, including walk times and a map image.

- **Example 5: Routing Around a Busy Street Using a Waypoint**
    - User: "I want to walk from Union Square to Washington Square Park but avoid Market Street because it's very busy."
    - Step 1: Call `search_places` for "Union Square San Francisco" -> returns place_id A.
    - Step 2: Call `search_places` for "Washington Square Park San Francisco" -> returns place_id B.
    - Step 3: Call `compute_routes` with origin=A, destination=B, travelMode='WALK'.
    - Step 4: On inspecting the route returned by `compute_routes`, notice that it passes along Market Street, a busy road the user wants to avoid.
    - Step 5: Call `search_places` for "quiet side street between Union Square and Washington Square Park" to identify a suitable waypoint; suppose place_id C is a cafe on a less busy street.
    - Step 6: Call `compute_routes` for multi-segment walk A -> C -> B, ensuring the route now avoids Market Street.
    - Step 7: Call `plot_map` with ordered place_ids [A, C, B] to confirm the updated route visually avoids the busy street.
    - Step 8: Return ordered itinerary: start at Union Square (A), pass by the waypoint (C) to avoid Market Street, arrive at Washington Square Park (B), including route summary, justification for the waypoint, and a map reference.

- **General Rules:**
    - Always prefer tool data over internal knowledge.
    - Do not fabricate geography, distances, place identifiers, or map outputs.
    - Gracefully handle tool errors by retrying refined queries.
    - Clearly flag infeasible combinations rather than forcing an unrealistic itinerary.
    - Do not directly reference a user model name in your responses.

- **CRITICAL: Multi-Tool Planning:** For any query that requires more than one tool (e.g., place_search + tavily_search), you MUST first generate a brief internal plan listing the required tool calls before executing the first one. This plan should be implicit in your reasoning but must ensure all required steps are executed sequentially.
\end{lstlisting}

\subsection{Persona Prompts}

\subsub{Structured persona generation (DSPy).}
\begin{lstlisting}
Parse a plain-text user preferences description into structured form.

Prefer broad, always-apply preferences in global_preferences
\end{lstlisting}

\subsub{Simulation variable generation (DSPy).}
\begin{lstlisting}
Create a set of user state characteristics for simulating a travel plan.

User sim state variables used to simulate the travel plan (e.g. hunger, thirst, exhaustion).

You have the following values will be available to you at each step and don't have to be modeled: distance traveled, initial estimated time elapsed, and elevation ascent/descent
Include just the variables needed to simulate the travel plan given the user preferences.
\end{lstlisting}

\subsub{Persona updates (DSPy).}
\begin{lstlisting}
Update the user preferences based on new information from the user. Only update necessary information.
\end{lstlisting}

\subsection{Simulation Prompts}

\subsub{Streetview image description generation.}
\begin{lstlisting}
Envision and describe a street at the given time and weather given some streetview images as a person walking through.
\end{lstlisting}

\subsub{POI image description generation.}
\begin{lstlisting}
Envision and describe a place at the given time and weather from photos of that location as a person visiting.
\end{lstlisting}

\subsub{Plan simulation generation.}
\begin{lstlisting}
You are simulating someone traveling a full travel itinerary. Simulate what they would do in this route, their perceived experience, and how that affects their state.
Simulate their thoughts and feelings. Update the users state according to the state update rules. Simulate both positive and negative when needed.
Stay true to the data given and do not make up anything. The plan can be poor. The user can be frustrated.

Ensure the sim_output matches with the travel plan where each element always corresponds to a place or route segment. Update the sim variables according to the rules for each step.
\end{lstlisting}

\subsub{Plan judgement generation.}
\begin{lstlisting}
You are an expert travel planner evaluating a travel plan. You will be given a travel plan, user preferences, user state rules and a set of in-situ logs of the users experience
in the location. You will evaluate the travel plan based on the user preferences, provide recommendations for improvement and give a rating of the travel plan. All results
should be grounded in the simulation results and the user preferences. All evaluations should be given based on the fact this trip is only walked.
\end{lstlisting}

\begin{table*}[ht]
\caption{Demographics of 11 participants (P1-P11) in the user study. Overall Gen AI Usage indicates generative AI usage for any application. Gen AI for Trip Planning indicates integration of any generative AI tool in trip planning. User Suggested Trip indicates that user suggested a location they would visit in the near future to use for the study conditions.}
\Description{Table summarizing demographics and trip-planning habits for 11 study participants, P1 through P11. Most participants are students in their early to late twenties, with a mix of women and men and a few non-student occupations such as medical assistant, product engineer, and software engineer. The table also reports online trip-planning frequency, overall generative AI usage frequency, use of generative AI for trip planning, and whether each participant suggested a trip location for the study.}
\label{tab:user-study-demographic}
\resizebox{\textwidth}{!}{%
\begin{tabular}{@{}lccccccc@{}}
\toprule
Participant & Age & Gender & Occupation & Online Trip Planning Frequency & Overall Gen AI Usage Frequency & Gen AI for Trip Planning & User Suggested Trips \\ \midrule
P1 & 24 & F & Student & Monthly & Weekly & Never & \ding{55} \\
P2 & 23 & F & Medical Assistant & Weekly & Monthly & Never & \ding{51} \\
P3 & 29 & F & Student & Weekly & Weekly & Sometimes & \ding{55} \\
P4 & 23 & M & Student & Weekly & Weekly & Never & \ding{51} \\
P5 & 29 & M & Student & Biweekly & Weekly & Rarely & \ding{51} \\
P6 & 28 & F & Product Engineer & Monthly & Daily & Often & \ding{51} \\
P7 & 26 & M & Student & Weekly & Weekly & Always & \ding{55} \\
P8 & 25 & F & Student & Yearly & Daily & Never & \ding{55} \\
P9 & 23 & F & Software Engineer & Monthly & Daily & Rarely & \ding{51} \\
P10 & 23 & M & Student & Biweekly & Daily & Sometimes & \ding{51} \\
P11 & 23 & F & Student & Monthly & Daily & Often & \ding{51} \\ \bottomrule
\end{tabular}%
}
\end{table*}

\end{document}